\documentclass[a4paper,11pt]{article}
\usepackage{pos}
\usepackage{tikz}

\title{Topological Data Analysis of Abelian Magnetic Monopoles in Gauge Theories}
\ShortTitle{TDA of Abelian Magnetic Monopoles}

\author*[a]{Xavier Crean}
\author*[b]{Jeffrey Giansiracusa}
\author*[a,c]{Biagio Lucini}

\affiliation[a]{Department of Mathematics, School of Mathematics and
Computer Science, Faculty of Science and Engineering, 
Swansea University Bay Campus, Fabian Way, Swansea SA1 8EN}

\affiliation[b]{Department of Mathematical Sciences, Durham University, Upper Mountjoy Campus, Durham, DH1 3LE, UK}

\affiliation[c]{School of Mathematical Sciences, Queen Mary University of London, Mile End Road, London, E1 4NS, UK}

\emailAdd{2237451@swansea.ac.uk}
\emailAdd{jeffrey.giansiracusa@durham.ac.uk}
\emailAdd{b.lucini@swansea.ac.uk}

\abstract{
Motivated by recent literature on the possible existence of a second higher-temperature phase transition in Quantum Chromodynamics, we revisit the proposal that colour confinement is related to the dynamics of magnetic monopoles using methods of Topological Data Analysis, which provides a mathematically rigorous characterisation of topological properties of quantities defined on a lattice. After introducing persistent homology, one of the main tools in Topological Data Analysis, we shall discuss how this concept can be used to quantitatively analyse the behaviour of monopoles across the deconfinement phase transition. Our approach is first demonstrated for Compact $U(1)$ Lattice Gauge Theory, which is known to have a zero-temperature deconfinement phase transition driven by the restoration of the symmetry associated with the conservation of the magnetic charge. For this system, we perform a finite-size scaling analysis of observables capturing the homology of magnetic current loops, showing that the expected value of the deconfinement critical coupling is reproduced by our analysis. We then extend our method to $SU(3)$ gauge theory, in which Abelian magnetic monopoles are identified after projection in the Maximal Abelian Gauge. A finite-size scaling of our homological observables of Abelian magnetic current loops at temporal size $N_t = 4$ provides the expected value of the critical coupling with an accuracy that is generally higher than that obtained with conventional thermodynamic approaches at comparable statistics, hinting towards the relevance of topological properties of monopole currents for confinement. 
}

\FullConference{The 41st International Symposium on Lattice Field Theory (LATTICE2024)\\
 28 July - 3 August 2024\\
Liverpool, UK\\}

\begin{document}

\maketitle

\section{Introduction}
\noindent
The phase structure of Quantum Chromodynamics (QCD) continues to be the focus of intense scrutiny (for a recent review, see, e.g., Ref.~\cite{Aarts:2023vsf}). The conventional picture is that at low temperatures quarks and gluons are confined into hadrons, while at higher temperatures a plasma of quarks and gluons characterises the physics at equilibrium. Recently, it has been suggested that a third regime opens between the latter two. This new {\em stringy fluid} regime is characterised by partial deconfinement (see Ref.~\cite{Glozman:2024xll} and references therein for an analysis based on emerging symmetries, and Ref.~\cite{Hanada:2023krw} for an approach based on the Polyakov loop).

Two quantities that have been used prominently in the lattice literature to characterise the phase structure of QCD are the chiral condensate and the Polyakov loop. However, neither is a strict order parameter in the presence of quarks with a finite mass. Significant progress could be made if the dynamics responsibile for colour confinement were exposed, since this would enable us to build {\em bona fide} order parameters. An intriguing suggestion is that the relevant degrees of freedom are of a topological nature. Appealing proposals identify those topological objects with field configurations carrying a non-trivial centre charge~\cite{tHooft:1977nqb} or a non-zero (Abelian) magnetic charge~\cite{tHooft:1981bkw}. Attempts to characterise  the phase structure of QCD and Yang-Mills theories in terms of symmetries of a topological nature have led to the construction of disorder parameters detecting the condensation of Abelian magnetic monopoles~\cite{DiGiacomo:1999fb,DiGiacomo:1999yas,Carmona:2001ja,Carmona:2002ty,Carmona:2002ye,DElia:2005sfk} and center vortices~\cite{DelDebbio:2000cx,DelDebbio:2000cb}. In the case of Abelian monopoles, these observables have been found to be dominated by lattice artefacts~\cite{Greensite:2008ss}. A potential issue that could have led to this non-negligible coupling with lattice artefacts is the reliance on semiclassical intuition for the construction of the dual order parameters. 

To shed more light on the problem of colour confinement in connection with the topology of gauge fields, in this work we employ Topological Data Analysis (TDA). TDA is a powerful and flexible data analysis toolset that provides robust computational methods for extracting and quantifying non-local topological features in data. TDA has a broad spectrum of applications. Recently, TDA approaches have been devised for specific investigations in Lattice Field Theories and Statistical Mechanics (for a non-exhaustive list of relevant studies, see Refs.~\cite{PhysRevResearch.2.043308,PhysRevB.104.104426,PhysRevD.107.034501,PhysRevB.104.235146,PhysRevE.105.024121,PhysRevB.106.085111,hirakida2020persistent,PhysRevD.108.056016,crean2024tda}). 

In this contribution, we shall use TDA methods to construct observables that capture the features of Abelian magnetic monopoles. After briefly introducing TDA from the perspective of our investigation (in Sect.~\ref{sec:essential-tda-for-lgt}), we define the relevant observables and study their behaviour across the deconfinement transition in Compact U(1) Lattice Gauge Theory in Sect.~\ref{sect:u1}. In Sect.~\ref{sect:su3}, this approach is then extended to monopoles in the Maximal Abelian Gauge in $SU(3)$ Yang-Mills, for which a study of the TDA observables is performed across the deconfinement phase transition.  Finally, we summarise our findings and possible future directions in Sect.~\ref{sect:conclusions}. 

\section{Essential Topological Data Analysis for Lattice  Gauge Theory}\label{sec:essential-tda-for-lgt}
\noindent

A lattice gauge theory is a special case of a general setup: we have a probability distribution $P$ on a complicated and high-dimensional configuration space $X$.  The distribution will depend on parameters such as a coupling constant or temperature, and we are interested in how $P$ changes with the parameters, with particular attention on phase transitions.   Our point in this paper is that TDA has a useful role to play in this.

TDA is a set of computational tools that provide numerical summaries of the shape of geometric objects.  There are two primary ways it can be used in studying statistical systems.  
\textit{Methodology A}: The geometric object that is fed to TDA could be a subspace of $X$ constructed from a set of samples, (e.g., the region where the density of $P$ is above some threshold), and TDA then tells us something about the shape of $P$. \textit{Methodology B:}  Each configuration might itself be a geometric object, or have content that can be represented by a geometric object, and so TDA-based invariants of these provide an interesting and useful class of observables through which the statistical behaviour of the system can be studied.  We explain each of these approaches in more detail below.

\subsection{Methodology A: the shape of a probability distribution}
The \emph{topology hypothesis} says that phase transitions of a system are closely related to changes in the topology of energy level sets of the configuration space, as explained in \cite{topology-hypothesis1, topology-hypothesis2, topology-hypothesis-theorem}.  TDA methods can allow one to capture a signal from these topology changes, as in \cite{topology-hypothesis-with-PH}.  We can use Markov Chain Monte Carlo (MCMC) to draw a large number of samples $x_1, x_2, \dots \in X$ from the distribution $P$.  A finite set of points in $X$ will of course be discrete, so there is not yet any interesting topology, but if we have a way of quantifying how similar two configurations are (a metric on $X$), then we can start to fill in some connective tissue between points that are close to one another to build something more interesting.  This process is a generalisation of reconstructing a circle from a finite number of points distributed around the circle by joining neighbours. The basic methods for accomplishing this are the Vietoris-Rips complex (described below in Sect.~\ref{sec:complexes}) and the $\alpha$-complex. When applied to the sampled points, this produces a model approximating the topology of the region where the distribution $P$ is sufficiently dense.  Topological invariants of this model then encode information about the shape of the distribution.  

\subsection{\label{sec:methodology-B}Methodology B: TDA-based observables}

In this mode, which is the one employed in our work, we first choose an algorithm $\mathscr{A}$ that takes a configuration $x\in X$ as input and outputs a geometric object.  TDA then provides numerical invariants of these geometric objects that describe their topology, and one examines the distribution of their values.  The most common invariants are \emph{homology} and \emph{persistent homology}, which we explain below.  The pipeline is: 
(1) Sample points $x_1, x_2, \ldots$ from $X$.
(2) Use the chosen algorithm to construct a geometric object $\mathscr{A}(x_i)$ for each sample.
(3) Compute topological invariants of each of these objects.
(4) Do statistical analysis with these invariants.
This methodology has been employed in~\cite{PhysRevE.98.012318, PhysRevE.103.052127, sym14091783, TDA-observable4, PhysRevResearch.2.043308, PhysRevB.104.104426}. 

\begin{figure}
\begin{center}
\includegraphics[scale=0.35]{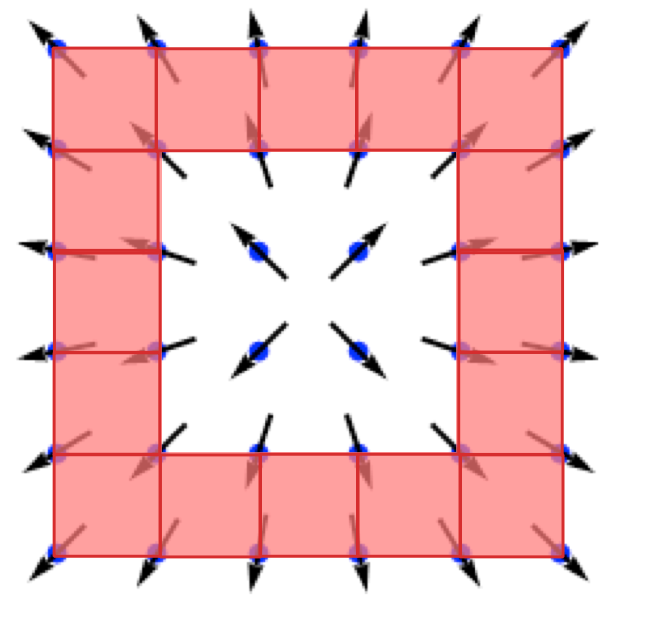}
\end{center}
\caption{\label{fig:vortex}In the 2d XY model, filling in plaquettes whose corners are nearly aligned will leave a hole around each (anti-)vortex.}
\end{figure}

To illustrate this with an example (from \cite{PhysRevE.105.024121}), let us consider the the 2d XY-model on an $n\times n$ lattice with periodic boundary conditions.  A configuration of this system consists of an element of the circle $S^1$ at each lattice site, and so the configuration space $X$ is a torus of dimension $n^2$.  The relevant topological structures in this system are vortices, so we might want to choose $\mathscr{A}$ to illuminate the vortices in a configuration.  The lattice partitions the 2d space on which the model lives into a toroidal chess board consisting of $n^2$ vertices, $2n^2$ edges, and $n^2$ plaquettes.  We build a subcomplex of this chessboard as follows.  We choose a threshold $\epsilon \in (0,\pi)$ and then fill in each edge if the spins at its two ends are aligned to within $\epsilon$.  For lack of anything better to do with the plaquettes, we fill in a plaquette if all 4 of its edges are filled in.  As in Fig.~\ref{fig:vortex}, this will tend to fill in everything except for a hole around the location of each vortex or anti-vortex.  Counting holes with homology (see below in Sect.~\ref{sec:homology}) then amounts to counting vortices!

\subsection{Simplicial complexes, cubical complexes, and filtrations}\label{sec:complexes}

Methodology A requires a way of approximately reconstructing a shape from a finite set of samples.  There are a few methods for achieving this, such as the \v{C}ech complex and the $\alpha$-complex, but the most common is the \emph{Vietoris-Rips construction}. It produces a \emph{simplicial complex}, which is a set of vertices together with the data of which collections of vertices span a simplex.  Two vertices might be connected by an edge, 3 might span a triangle, and so on. Simplicial complexes are a standard data structure in computational topology software, such as GUDHI \cite{gudhi}.

\begin{figure}
\begin{center}
\begin{tikzpicture}
  \coordinate (A) at (0,0);
  \coordinate (B) at (1.2,0.2);
  \coordinate (C) at (0.8,1.4);
  \coordinate (D) at (-0.8,1.2);

  \draw[blue] (A) circle (1);
  \draw[blue] (B) circle (1);
  \draw[blue] (C) circle (1);
  \draw[blue] (D) circle (1);

  \fill[red!30, opacity=0.3] (A) -- (B) -- (C) -- cycle;
  \fill[red!30, opacity=0.3] (A) -- (B) -- (D) -- cycle;
  \fill[red!30, opacity=0.3] (A) -- (C) -- (D) -- cycle;
  \fill[red!30, opacity=0.3] (B) -- (C) -- (D) -- cycle;

  \draw[red] (A) -- (B);
  \draw[red] (A) -- (C);
  \draw[red] (A) -- (D);
  \draw[red] (B) -- (C);
  \draw[red] (C) -- (D);

  \fill[black] (A) circle (2pt);
  \fill[black] (B) circle (2pt);
  \fill[black] (C) circle (2pt);
  \fill[black] (D) circle (2pt);

  \begin{scope}[shift={(4,0)}]
    \coordinate (A2) at (0,0);
    \coordinate (B2) at (1.2,0.2);

    \draw[blue] (A2) circle (1);
    \draw[blue] (B2) circle (1);

    \draw[red] (A2) -- (B2);

    \fill[black] (A2) circle (2pt);
    \fill[black] (B2) circle (2pt);
  \end{scope}

\end{tikzpicture}

\end{center}
\caption{The Vietoris-Rips simplicial complex built from 6 points in the plane.\label{fig:vietorisrips}} 
\end{figure}
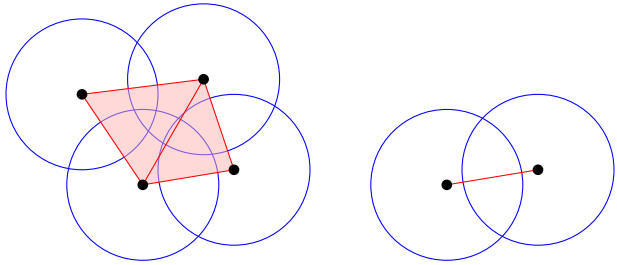

The set of vertices is simply our set of sampled points, and then a finite subset $\{y_0, ... y_d\}$ spans a $d$-simplex if the distance between each pair is smaller than some chosen threshold $t$.  This is roughly equivalent to placing a ball of radius $r$ around each sampled point and then taking the union of these balls (Fig.~\ref{fig:vietorisrips}).  The Vietoris-Rips complex $VR_r$ is then a model approximating the topology of the region where the distribution $P$ is sufficiently dense.

One can fix a single value of $r$, but it is often useful to think about how the topology evolves as $r$ sweeps across a range.  Increasing $r$ can add some additional simplices, but it clearly cannot take any away, so as $r$ increases we have an increasing sequence of simplicial complexes  $VR_{r_1} \subset VR_{r_2} \subset \cdots$.  Such a nested sequence of subspaces is called a \emph{filtration}.  The idea of \emph{persistence} is to examine the homology of these complexes along the filtration.

When applying Methodology B to systems defined on a cubical lattice, the most natural kind of geometric object is instead a \emph{cubical complex} (often equipped with a filtration).  For a $d$-dimensional system, the lattice partitions the underlying space into vertices (the lattice sites), edges, plaquettes, cubes, hypercubes up to dimension $d$.  We will refer to each of these components as a cell.
A cubical complex is simply a subset of these cells subject to the following condition: if a cell $C$ is present, then all lower dimensional cells contained in its boundary are also present.  (More general notions of cubical complex, analogous to the definition of a simplicial complex above, can be found in the literature, but we will not need them.)

\subsection{Counting holes with homology}\label{sec:homology}

Holes come in different types.  A point missing from the plane is a 1-dimensional hole in the sense that it can be captured by a circle, which is a 1-dimensional manifold.  However, the circle cannot capture  a point missing from 3-space because it could always slip off over the top or bottom; this hole must be wrapped by a sphere, so it is a 2-dimensional hole.  \emph{Homology} is a tool from algebraic topology that formalises the idea of counting the holes and voids.  The input is a simplicial or cubical complex $X$, and for each natural number $n$ there is a vector space $H_n(X)$.  The dimension of this vector space represents the count of $n$-dimensional holes.  

Here are some useful properties of homology.
\begin{enumerate}
\item If $X$ is a point (or contractible), then $H_0(X)$ is 1-dimensional, and $H_n(X) = \{0\}$ for all $n>0$.
\item In general, $H_0(X)$ has dimension equal to the number of connected components of $X$.  
\item Disjoint union $X\sqcup Y$ corresponds to direct sum of homology.
\item If $n$ is larger than the dimension of $X$, then $H_n(X)=\{0\}$. 
\item If $X$ is an $n$-dimensional closed and oriented manifold, then $H_n(X)$ is 1-dimensional.
\end{enumerate}

Implicitly, we have chosen a coefficient field to work with, so the homology is a vector space over this field.  In practice, one often uses $\mathbb{Z}/2$.  Note that, with $\mathbb{Z}/2$ coefficients, all closed manifolds behave as if they are oriented.

The fact that homology outputs a vector space rather than just a number is extremely important because a continuous map $X\to Y$ induces a linear map $H_n(X) \to H_n(Y)$.  This is essential in defining persistent homology.

\begin{figure}
\begin{center}
\begin{tikzpicture}[scale=0.7]

  \shade[inner color=darkgray, middle color=darkgray!10, outer color=white, shading angle=45] (-0.5,-0.5) rectangle (4.5,3.5);
  \fill[white] (2,1) rectangle (3,2);

middle color=lightgray!50
  \draw[gray] (2,1) rectangle (3,2);

  \draw[blue, thick] (1,0) rectangle (4,3);

  \foreach \x in {1,2,3,4} {
    \fill[black] (\x,0) circle (1.5pt);
    \fill[black] (\x,3) circle (1.5pt);
  }
  \foreach \y in {0,1,2,3} {
    \fill[black] (1,\y) circle (1.5pt);
    \fill[black] (4,\y) circle (1.5pt);
  }

  \fill[green,opacity=0.3] (0,1) rectangle (0.9,3);
  \draw[gray,thick] (0,1) rectangle (0.9,3);
  \foreach \y in {1,2,3} {
    \fill[black] (0,\y) circle (1.5pt);
  }

 \draw[red, very thick] (-0.1,0.9) -- (-0.1,3.1) -- (4.1,3.1) --
 (4.1,-0.1) -- (0.9,-0.1) -- (0.9, 0.9) --  (-0.1,0.9);

  \node at (0.5,2) {$w$};
  \node[text=red] at (0.5,0.5) {$z'$};
  \node[text=blue] at (1.5,0.5) {$z$};

  \node at (5.5,2) {$z - z' = \partial w$};
\end{tikzpicture}
\end{center}
\caption{The blue edges form a 1-cycle $z$, and the red edges form a 1-cycle $z'$.  The difference $z-z'$ is the boundary of the sum $w$ of the two plaquettes in green.\label{fig:cycle}}
\end{figure}
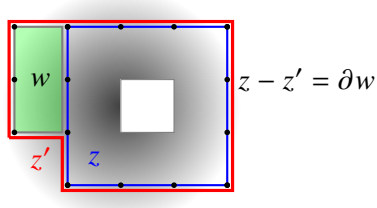

The computation of homology can be performed algorithmically.  Given a simplicial or cubical complex $X$ with a chosen orientation for each cell in $X$, let $C_i$ be the vector space with basis given by the set of $i$-dimensional cells in $X$.  The boundary operator is the linear map $\partial_i: C_i \to C_{i-1}$ defined by sending each $i$-cell to the sum of its faces with signs determined by the orientations.   A \emph{cycle} is an element $z\in C_i$ such that $\partial(z) = 0$.  The homology $H_i(X)$ is then defined to be the quotient of the space of cycles where two elements $z, z'$ are identified if the difference $z-z'$ is in the image of $\partial_{i+1}$ (see Fig.~\ref{fig:cycle}).  Computing the dimension or finding a basis for the homology is accomplished by standard algorithms in linear algebra.

Now suppose $X$ is a 1-dimensional (simplicial or cubical) complex, so it is a graph.  In this case the homology is particularly easy to describe.  If $X$ is connected, then by contracting a spanning tree, one sees that any connected graph is homotopy equivalent to a collection of circles joined at a single point (one circle for each edge not contained in the spanning tree). Hence, for a general graph $X$, $\dim H_0(X)$ is the number of connected components, and  $\dim H_1(X)$ is the number of circles. 

\subsection{Persistent homology}

Persistent homology is an enhancement of homology.  The input is a geometric object equipped with a filtration , $X_1 \subset X_2 \subset \cdots$, and the output is a representation of how the homology $H_n(X_i)$ varies along this sequence. 

An important property of homology is that a continuous map $X \to Y$ induces a linear map $H_i(X) \to H_i(Y)$.  Hence, when we feed it a filtration, we get a sequence of vector spaces and linear maps $H_i(X_1) \to H_i(X_2) \to H_i(X_3) \to \cdots$.  One can always choose a basis for a vector space, but for a sequence of vector spaces we have a less trivial fundamental fact from linear algebra.

\noindent\textbf{Proposition.}\textit{ Given a sequence of finite dimensional vector spaces $V_n$ connected by linear maps $V_n \to V_{n+1}$, it is always possible to choose bases for the $V_n$ that are compatible with the linear maps in the following sense:  the image of each basis vector in $V_n$ is either zero or a basis vector of $V_{n+1}$, and each basis vector of $V_n$ is the image of at most a single basis vector in $V_{n-1}$. }

This means that the basis vectors link up into chains, with each chain starting in some $V_i$ and ending at some $V_j$; the index $i$ is the \emph{birth time}, and the index $j$ is the \emph{death time} for this chain.  The collection of these (birth, death) pairs completely determines the isomorphism class of the sequence of vector spaces.  When applied to the homology of a filtration of a geometric object $X$, this invariant is called the \emph{persistent homology} of $X$.

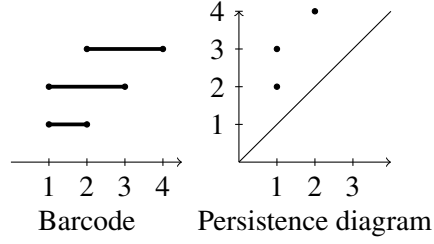
\begin{figure}
\begin{center}
\begin{tikzpicture}[scale=0.5]
    \draw[->] (0,0) -- (4.5,0) node[right] {};
    \foreach \x in {1,2,3,4} {
        \draw (\x,0.1) -- (\x,-0.1) node[below] {\x};
    }
    \draw[line width=0.5mm] (1,1) -- (2,1);
    \filldraw (1,1) circle (2pt);
    \filldraw (2,1) circle (2pt);
    \draw[line width=0.5mm] (1,2) -- (3,2);
    \filldraw (1,2) circle (2pt);
    \filldraw (3,2) circle (2pt);
    \draw[line width=0.5mm] (2,3) -- (4,3);
    \filldraw (2,3) circle (2pt);
    \filldraw (4,3) circle (2pt);
    \node[below] at (2,-1) {Barcode};

    \draw[->] (6,0) -- (10,0) node[right] {};
    \draw[->] (6,0) -- (6,4) node[above] {};

\foreach \x in {7,8,9} {
    \draw (\x,0.1) -- (\x,-0.1) node[below] {\the\numexpr\x-6\relax};
}

     \foreach \y in {1,2,3,4} {
        \draw (6.1,\y) -- (5.9,\y) node[left] {\y};
    }
    \draw (6,0) -- (10,4);
    \filldraw (7,2) circle (2pt);
    \filldraw (7,3) circle (2pt);
    \filldraw (8,4) circle (2pt);
    \node[below] at (8,-1) {Persistence diagram};
\end{tikzpicture}
\end{center}
\caption{A barcode representation of persistent homology on the left, and a persistence diagram representation of the same peristent homology on the right.\label{fig:persrep}}
\end{figure}

There are two commonly used graphical representations for persistent homology (Fig.~\ref{fig:persrep}).  
\begin{itemize}
\item \textbf{Barcodes:}  Each (birth, death) is depicted as an interval from the birth time to the death time, so a barcode is a multiset of intervals.
\item \textbf{Persistence diagrams:}  Each (birth, death) pair is depicted as a point in the plane above the diagonal.
\end{itemize}

\subsection{Stability}
There is a metric on the space of persistence diagrams called the \emph{bottleneck metric}. The \emph{Stability Theorem} (e.g. \cite{Bauer-Lesnick}) guarantees that a small perturbation of the entry times of cells in a filtration leads to a corresponding small change in the resulting persistence diagram with respect to this metric.   Note that a small perturbation of the positions of the points from which one builds a Vietoris-Rips complex leads to such a filtration perturbation.

\subsection{Vectorising persistence diagrams}

The set of persistence diagrams (or barcodes) does not form a vector space, and this is problematic if we want to do statistical analysis and machine learning on a distribution of persistence diagrams.  For example, it is not clear what the sample mean of a collection of persistence diagrams is.  To deal with this obstacle, one usually employs a map from the space of persistence diagrams to a finite dimensional vector space.  This is called \emph{vectorisation}; there are by now a multitude of methods, and they are surveyed in \cite{Nanda-vectorisation-survey}.  

One of the most intuitively clear methods is \emph{persistence images}, \cite{Adams-persistence-images}. The idea is to first take a persistence diagram and turn it into a smooth function by replacing each point with a small Gaussian, and then turn this function into a grayscale image by setting each pixel value to be the integral of the function over the corresponding square.


\section{Topological data analysis of monopole current networks in compact $U(1)$ Lattice Gauge Theory}
\label{sect:u1}
\noindent
In order to develop methods sensitive to topological defects in full QCD, our first step has been to analyse a simple, well-understood model involving magnetic monopole current networks. We find that these can be analysed across the deconfinement phase boundary using methods from Sect.~\ref{sec:essential-tda-for-lgt}.

\subsection{$4$-dimensional $U(1)$ Lattice Gauge Theory at zero temperature}
\noindent
We consider $4$-dimensional compact $U(1)$ Lattice Gauge Theory at zero temperature on  a cubical lattice $\Lambda = \{0,...,L-1\}^{4}$ with periodic boundary conditions. The lattice $\Lambda$ is a discrete representation of the $4$-torus $T^{4} = S^{1} \times S^{1} \times S^{1} \times S^{1}$. An edge $x \to x + \hat{\mu}$ (referred to as a link) is indexed by $(x \in \Lambda, \mu \in \{0,1,2,3\})$. The gauge field is represented by link variables $U_{\mu}(x) = e^{i \theta_{\mu}(x)} \in U(1)$ such that $\theta_{\mu}(x) \in (-\pi, \pi]$. Given the plaquette indexed by $(x \in \Lambda, \mu \nu)$, the value of the Wilson loop around this plaquette is $U_{\mu\nu}(x) = e^{i \theta_{\mu\nu}(x)}$, where we have the oriented sum
\begin{equation}\label{eq:plaquette-phase}
    \theta_{\mu\nu} (x) \equiv \theta_{\mu}(x) + \theta_{\nu}(x + \hat{\mu}) - \theta_{\mu}(x + \hat{\nu}) - \theta_{\nu}(x) \quad \in (-4\pi, 4\pi].
\end{equation}
This plaquette phase $\theta_{\mu\nu} (x)$ is related to the field strength tensor via $\theta_{\mu\nu} (x) = a^{2} F_{\mu\nu}(x) + O(a^{3})$, where $a$ is lattice spacing, and, by relation $B_{i}(x) = \frac{1}{2} \varepsilon_{ijk} F_{jk}(x) $, it represents the amount of magnetic flux through the plaquette. We take the standard Wilson action 
\begin{equation}\label{eq:wilson-action}
    S = \beta \sum_{x} \sum_{\mu < \nu} ( 1 - \cos[\theta_{\mu\nu}(x)] ),
\end{equation}
with inverse coupling $\beta \equiv \frac{1}{g^{2}}$, which converges to the Maxwell action in the classical continuum limit $a \to 0$.

\subsection{Magnetic monopole currents}
\noindent
On the $d=3$ lattice $\Lambda$, since a plaquette $\theta_{\mu\nu} (x)$ represents the smallest area to carry magnetic flux, a unit $3$-cube, bounded by six surface plaquettes, represents the smallest volume to carry magnetic charge. We interpret a point charge to live at the centre of a $3$-cube or, conversely, on a vertex of the dual lattice $\Lambda^{*}$. In $d=4$, point charges sweep-out current lines such that we have a dual picture where magnetic currents live on links of the $4$-dimensional dual lattice $\Lambda^{*}$. In our convention, a current on link $(x \in \Lambda^{*},\rho)$ is associated to a $3$-cube at lattice site $x + \hat{\rho} \in \Lambda$. For $x \in \Lambda^{*}$, magnetic current is specified by the relation
\begin{align}\label{eq:magnetic-current}
    j_{\rho} (x) &= \Delta_{\sigma} \theta_{\rho \sigma}^{*} (x + \hat{\rho}),
\end{align}
where the dual plaquette phase is defined as $ \theta_{\rho \sigma}^{*} (x) 
 \equiv \frac{1}{2} \varepsilon_{\rho \sigma \mu \nu} \theta_{\mu \nu} (x)$ and $\Delta_{\sigma}$ is the forward finite difference operator. By the lattice equivalent of the Bianchi identity, the right hand side of Eq.~\eqref{eq:magnetic-current} is identically zero. Intuitively, Eq.~\eqref{eq:magnetic-current} is zero since it represents a sum of the magnetic flux through a closed surface (the six faces of a unit $3$-cube). 

 Following De Grand and Toussaint in Ref.~\cite{PhysRevD.22.2478}, we would like to detect magnetic monopoles which are sources/sinks in the magnetic field. Monopoles are defined in terms of unphysical gauge variant Dirac strings interpreted as infinitely thin solenoids. A Dirac string carries a unit of $2\pi$ flux through the surface of a plaquette. We may remove the Dirac string content from a plaquette phase $\theta_{\mu \nu} (x) \in (-4\pi, 4\pi]$ by taking the modulus with respect to $2\pi$ such that the remaining part $\bar{\theta}_{\mu\nu} (x) \in (-\pi, \pi]$ is the physical flux: $\theta_{\mu \nu} (x) = \bar{\theta}_{\mu\nu} (x) + 2\pi n_{\mu\nu} (x)$. The number $n_{\mu\nu}(x) \in \{-2, -1, 0, 1, 2\}$ counts the number of Dirac strings passing through a plaquette. By considering the physical flux only, we now see that Eq.~\eqref{eq:magnetic-current} can be non-zero when considering a monopole. Previously, when considering the closed surface of a unit $3$-cube, the outgoing magnetic flux was counterbalanced by the incoming Dirac string(s). However, whilst unphysical Dirac strings can be moved around by gauge transformation, the outgoing flux is gauge invariant and thus represents a physical source of the magnetic field interpreted as a Dirac monopole. The procedure for identifying monopole currents on $\Lambda^{*}$ becomes clear: $\forall \, x \in \Lambda$, one needs to compute the four monopole numbers associated to $3$-cubes $M_{\rho}(x) = -\frac{1}{2} \varepsilon_{\rho \sigma \mu \nu} \Delta_{\sigma} n_{\mu \nu} (x)$ and then identify this number with the corresponding link on $\Lambda^{*}$ via $j_{\rho}(x) = M_{\rho} (x + \hat{\rho})$.  Fixing an orientation, given link $(x \in \Lambda^{*}, \rho)$, we interpret $j_{\rho} (x) = 0$ as no current line on the link  and $j_{\rho} (x) = \pm 1 $ (or $\pm 2$) as one (or two) current line(s) on the link.

Monopole currents must obey current conservation $\Delta_{\rho}^{*} j_{\rho} (x) = 0$, where $\Delta_{\rho}^{*}$ is the backward finite difference operator, and so they form closed loops; we will leverage this information when computing topological invariants of monopole currents. We refer to a connected set of monopole current lines as a network. Since our lattice $\Lambda$ is a discrete model of $T^{4}$, monopole currents can either be local current loops that do not wrap in any of the directions on $T^{4}$ or global current loops that wrap in any or all of the periodic directions of $T^{4}$. Whilst the Bianchi identity of Eq.~\eqref{eq:magnetic-current} is violated locally in order to define magnetic monopoles, for the configuration as a whole, it must still hold. Since any wrapping current loop contributes a non-zero charge to the configuration, there must also exist an oppositely oriented partner loop. 

\subsection{The deconfinement phase transition}
\noindent
In $d=4$, there exists a weak first order phase transition that has been well studied in the literature -- see Refs.~\cite{Creutz:1979zg,Lautrup:1980xr,Bhanot:1981tn,Jersak:1983yz,Barber:1984ak,Evertz:1984pn,Barber:1985at,GROSCH1985171,Lang:1986yd,DelDebbio:1994sx,Lang:1994ri,Kerler:1994qc,Kerler:1995va,Kerler:1995nj,Jersak:1996mn,Jersak:1996mj,Kerler:1996sf,Kerler:1996cr,Campos:1997br,Campos:1998jp,Vettorazzo:2003fg,DiCairano:2023kzh} for an incomplete history of both numerical and theoretical developments -- using various specialised numerical methods, e.g., Refs.~\cite{Arnold_2001,langfeld2016efficient}. In the low-$\beta$ phase, a confining potential is caused by a magnetic monopole condensate; this has been studied in the context of a monopole-creation order parameter in Ref.~\cite{DelDebbio:1994sx}. In Ref.~\cite{Kerler:1995va}, studying a lattice with periodic boundary conditions, Kerler et al. were able to establish that the transition is of percolation-type. In the low-$\beta$ confining phase, there exists a percolating network of monopole currents that wrap in all four directions on $T^{4}$; whereas, in the high-$\beta$ deconfined phase, there do not exist any wrapping currents. The signature of the transition could be detected by looking at whether current networks wrapped or not. To analyse whether the periodic boundary conditions of the lattice were the cause of the phase transition, the model was studied (e.g., Ref.~\cite{Campos:1998jp}) on a space homeomorphic to the $4$-sphere $S^{4}$, where wrapping currents do not exist since any loop is contractible. The transition was still detectable indicating that monopole condensation is not a consequence of the periodic boundary conditions. 

\subsection{Homological observables for current networks}
\noindent
We leverage Methodology B from Sect.~\ref{sec:methodology-B} to construct observables that characterise the structures formed by monopole current networks. Not only do these observables precisely reveal the signature of the deconfinement transition but they provide an interpretable geometric picture of the current networks. This topological characterisation of currents does not involve looking for wrapping networks on $T^{4}$ and so we expect that our analysis will be applicable in the setting of a lattice discretisation of $S^{4}$. For details on our computational pipeline, we point the reader to Refs.~\cite{crean2024tda,ZENODO2}.

Feom the monopole currents we produce a graph $X_{j}$. We choose arbitrary orientations for the edges in order to define the chain complex from which homology is computed (but the resulting homology is independent of these choices).
We then compute $b_{0}(X_{j}) \equiv \dim H_{0}(X_{j})$, the number of connected components of $X_{j}$, and $b_{1}(X_{j}) \equiv \dim H_{1}(X_{j})$, the number of loops of $X_{j}$, and then estimate their expectation value by the mean of $N$ sample configurations. To compare across lattice sizes, we normalise by the volume $V=L^{4}$:
\begin{equation}\label{eq:betti-number-observables}
    \rho_{0} \equiv \langle b_{0} \rangle / V,
    \quad
    \rho_{1} \equiv \langle b_{1} \rangle / V.
\end{equation}
Further, we also compute the respective susceptibilities
\begin{equation}\label{eq:betti-number-susceptibilities}
    \chi_{0} \equiv (\langle b_{0}^{2} \rangle -\langle b_{0} \rangle^{2}) / V,
    \quad
    \chi_{1} \equiv (\langle b_{1}^{2} \rangle -\langle b_{1} \rangle^{2}) / V.
\end{equation}
In Sect. \ref{sec:numerical-results}, we use these observables to estimate the inverse critical coupling $\beta_{c}$ in the infinite volume limit via a finite-size scaling analysis.

\subsection{Numerical set-up and results}\label{sec:numerical-results}
\noindent
We generate $N=200$ sample configurations via MCMC simulation where each composite sweep consists of $1$ approximate heat-bath update~\cite{Bazavov:2005zy} and $5$ over-relax updates. In the critical region, as the lattice size $L$ increases, the simulation suffers from a very long auto-correlation time caused by meta-stabilities formed by wrapping current networks. Therefore the probability of tunnelling between phases is exponentially suppressed, which can introduce large systematic error. To mitigate these effects, we measure samples infrequently -- at least every $42,000$ composite sweeps. In this respect, we are limited by our computational resources; however, in this study, our objective was not necessarily to achieve high precision (in the same vein as other specialised numerical studies using a large number of statistics, e.g., Refs.~\cite{Arnold_2001,langfeld2016efficient}) but to design a robust computational pipeline that gives us precise and reliable results with a modest number of statistics.

\begin{figure}
\begin{center}
\includegraphics[width=0.45\textwidth]{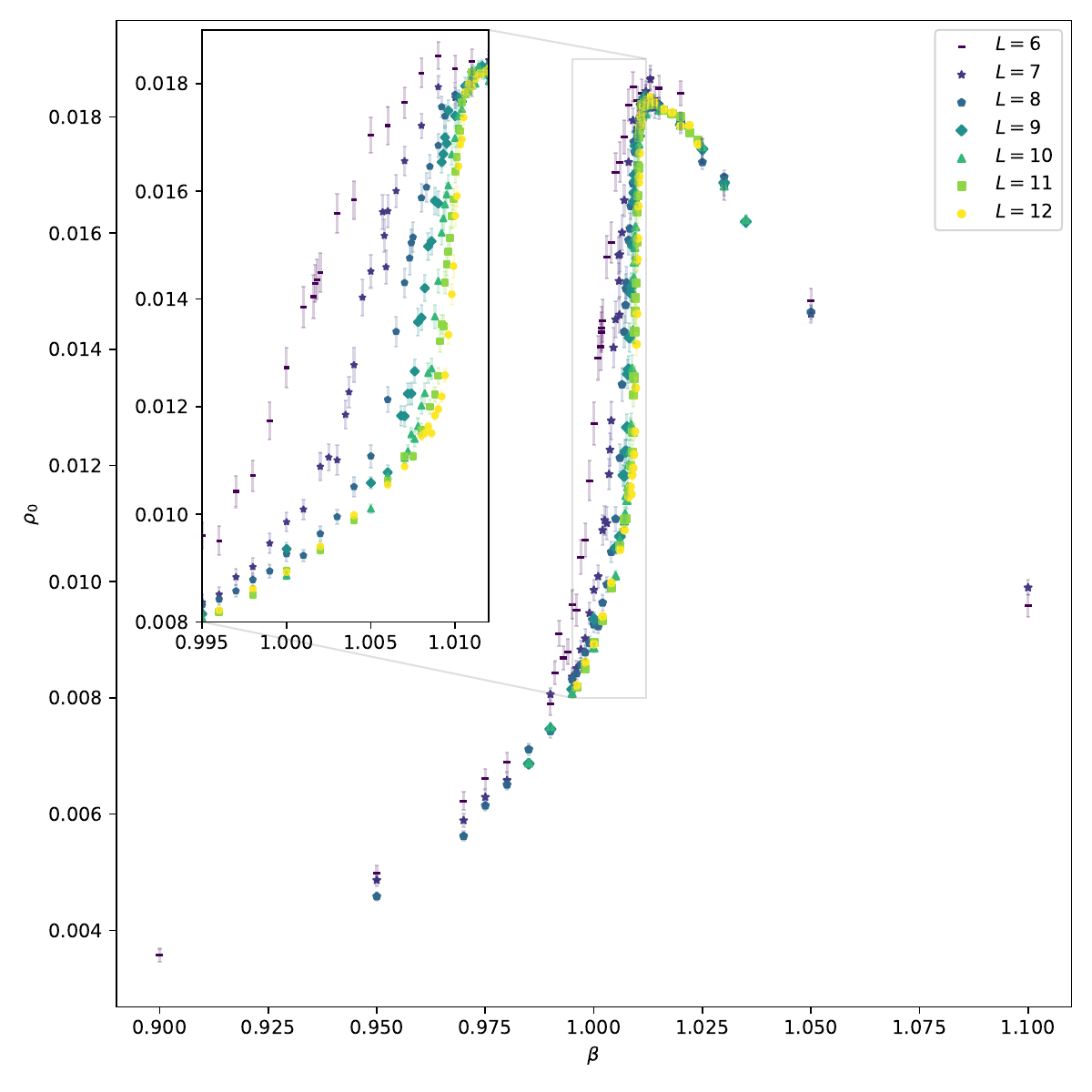}
\hspace{0.05\textwidth}
\includegraphics[width=0.45\textwidth]{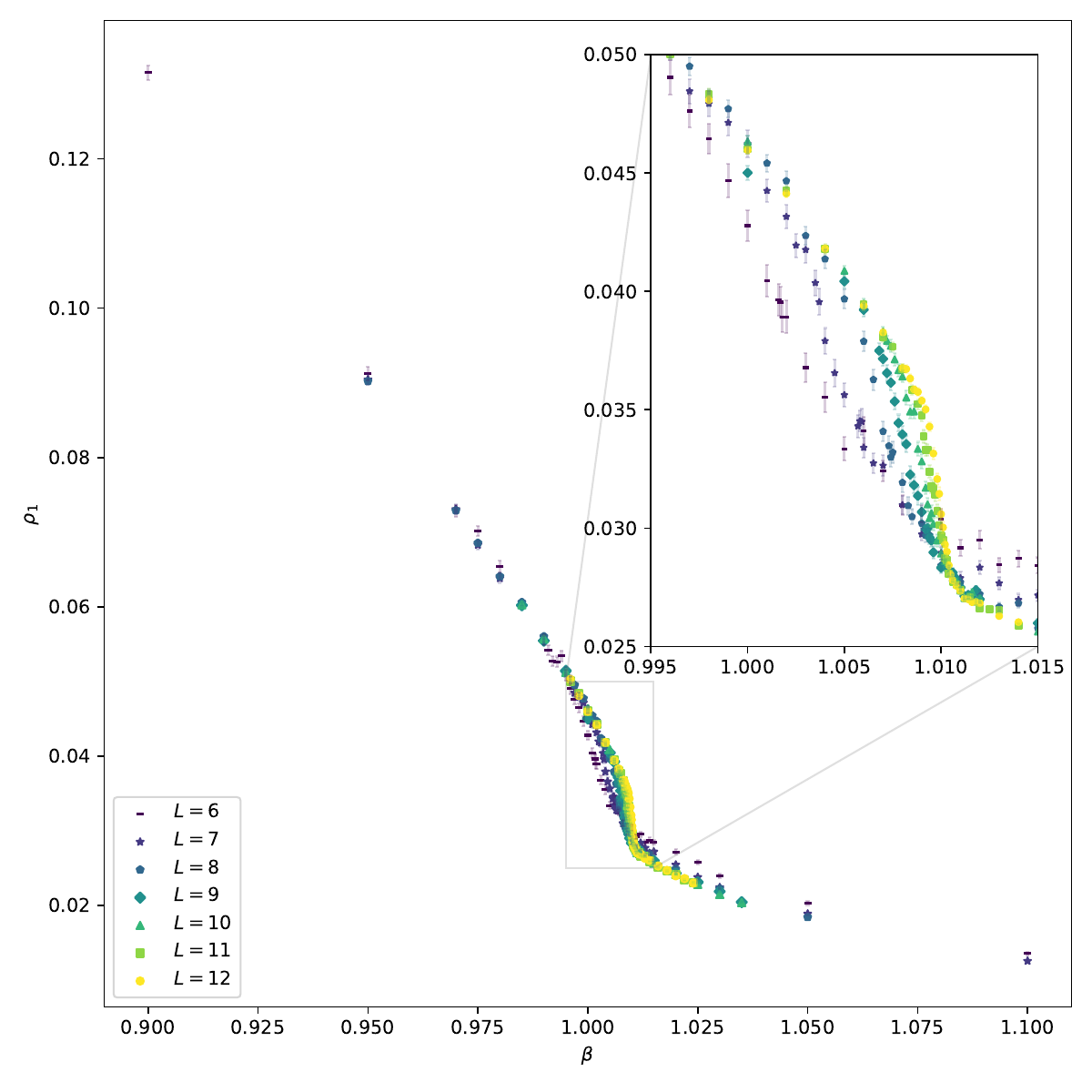}
\end{center}
\caption{The behaviour of the Betti number observables $\rho_0$ (left) and $\rho_1$ (right) across the deconfinement transition for a range of lattice sizes $L$ as indicated. Inset plots are zoom-ins of the critical region. Error bars are computed by bootstrapping with $N_{\text{bs}} = 500$.}
\label{fig:sect3:rho0_rho1}
\end{figure}

We plot the Betti number observables $\rho_{0}$ and $\rho_{1}$ in Fig.~\ref{fig:sect3:rho0_rho1}. The behaviour exhibited by $\rho_{0}$ and $\rho_{1}$ aligns with the accepted physical picture seen in the literature (e.g., Ref.~\cite{Kerler:1994qc}). In the low-$\beta$ phase, we see a small $\rho_{0}$ corresponding to a small number of connected components and a large $\rho_{1}$ corresponding to large number of loops consistent with a percolating current network. Whereas, in the high-$\beta$ phase, we see a decreasing $\rho_{0} \approx \rho_{1}$ consistent with independent small current networks that roughly have the same number of components as loops. Moreover, the susceptibilities $\chi_{0}$ and $\chi_{1}$ peak in the critical region, thus allowing us to extract the pseudo-critical $\beta_{c}(L)$ for a range of lattice sizes $L=6,7,...,12$. We may extract a more precise estimate of the locations of the peak of the respective susceptibilities via histogram reweighting (see Ref.~\cite{ferrenberg1989optimized}) which may then be extrapolated to the infinite volume limit $L\to \infty$ via a finite-size scaling analysis using the ansatz
\begin{equation}\label{eq:sec3-ansatz}
    \beta_{c} (L) = \beta_c + \sum_{k=1}^{k_{\text{max}}} B_{k} V^{-k}.
\end{equation}
We find that our results in Tab.~\ref{tab:sec3-FFS-beta_c}, for estimating the critical inverse coupling of the transition $\beta_c$, are consistent with a standard observable used to probe this transition, i.e., the average plaquette action $E = \frac{1}{6V} \sum_{x , \mu < \nu} \cos\theta_{\mu\nu}(x)$. 
\begin{table}[]
    \centering
    \begin{tabular}{|c|c|c|}
        \hline
        $E$ & $\rho_{0}$ & $\rho_{1}$ \\
        \hline
         $1.01071(3)$ & $1.01076(6)$ & $1.01076(6)$ \\
         \hline
    \end{tabular}
    \caption{Estimates for the critical inverse coupling $\beta_{c}$, in the infinite volume limit, obtained via finite-size scaling analysis for the average plaquette action $E$ and homological observables $\rho_{0}$ and $\rho_{1}$.}
    \label{tab:sec3-FFS-beta_c}
\end{table}

\subsection{Persistent homology as an encoder of current geometry}
\noindent
We would like to use persistent homology to further analyse the structures formed by monopole currents. There are many choices of filtration and the challenge is finding a meaningful one that retains sufficient, interpretable information. Inspired by the Vietoris-Rips filtration that expands balls of radius $\varepsilon \in \mathbb{R}_{\geq 0}$ around points in a pointcloud and builds a simplicial complex $VR_\varepsilon$ based on the intersection of these balls (see Sect.~\ref{sec:essential-tda-for-lgt}), we design a filtration that expands (in units of the lattice spacing) a $4$-dimensional tubular volume radially outwards from current lines. Topological features will be born and die according to the geometry of the current lines in a configuration. This filtration characterises both the intra-network and inter-network structure of currents and thus retains more information than the homologies of the networks alone.\footnote{If, for example, a set of current networks lived on a closed $k$-dimensional submanifold, with $k=1,2,3$, a long-lived $k$-dimensional feature would be detected by this filtration.} Since there will be variation across sample configurations, the aim is to average across $N$ sample persistence diagrams to encode a numerical summary of the geometry of monopole current networks at a given $\beta$.

The numerical set-up is the same as in Sect.~\ref{sec:numerical-results} with $N=200$ sample configurations for each lattice size $L=6,7,...,12$ respectively. After computing the persistence diagrams, we find that there exists a significant difference in the number and spread of points in the low-$\beta$ and high-$\beta$ phases, which, we hypothesise, corresponds to the lower density of current loops in the high-$\beta$ regime -- the filtration has more space between currents to explore and thus a greater number of topological features are constructed by the filtration. In the high-$\beta$ phase, we observe that there are no statistically significant long-lived topological features. These results and the underlying methodology will be the subject of a forthcoming publication. Extensions to Abelian monopoles in Yang-Mills theories are also in progress.

\section{Topological data analysis of Abelian monopole current networks in $SU(3)$ Lattice Gauge Theory}\label{sect:su3}
As another step in the direction of studying the full QCD case, we generalise the homological observables used for the analysis of the deconfinement phase transition in compact $U(1)$ Lattice Gauge Theory to $SU(3)$ Yang-Mills theory on a lattice and provide a first set of results on the behaviour of these observables across the deconfinement phase transition of the latter theory. 

\subsection{The Lattice setup}
\label{subsec:TDASU3:lattice}
\noindent
In this work, we use the Wilson action of lattice $SU(3)$ Yang-Mills, which is given by
\begin{equation}
    S = \beta \sum_{i, \mu < \nu} \left( 1 - \frac{1}{3} {\cal R} \mathrm{e~Tr} U_{\mu \nu}(i) \right) \ , 
\end{equation}
where $\beta = 6/g_0^2$ ($g_0$ being the bare coupling), ${\cal R} \mathrm{e~Tr} U_{\mu \nu}(i)$ indicates the real part of the trace of the plaquette $U_{\mu \nu}(i)$, with the latter quantity being the ordered product of the link variables $U \in SU(3)$ along the elementary plaquette stemming from point $i$ and spanning the positive directions $\hat{\mu}$ and $\hat{\nu}$ of the lattice. More explicitly,
\begin{equation}
 U_{\mu \nu}(i) = U_{\mu}(i) U_{\nu}(i + \hat{\mu}) U^{\dag}_{\mu}(i + \hat{\nu}) U^{\dag}_{\nu}(i) \ , 
\end{equation}
with $U_{\mu}(i)$ being the link variable associated with the link stemming from $i$ in the positive $\hat{\mu}$ direction and $U^\dag_{\mu}(i)$ its complex conjugate. The system is formulated on an $S^1 \times T^3$ lattice, with the extension of the $S^1$ kept fixed and defining the temperature $T$ through the relationship
\begin{equation}
    T = (a N_t)^{-1} \ , 
\end{equation}
where $a$ is the $\beta$-dependent lattice spacing and $N_t$ is the number of lattice sites discretising the $S^1$. The corresponding direction is conventionally referred to as the temporal direction. The $T^3$ torus identifies three spatial directions that we take of identical length $N_s$. We shall take the $N_s \to \infty$ limit to consider the system in the thermodynamic limit. 

The path integral of the system 
\begin{equation}
Z = \int \left( {\cal D} U \right) e^{- S} 
\end{equation}
is sampled with MCMC methods using both the heat-bath and the overrelaxation algorithm in a ratio 1:4. We define an iteration of one heat-bath step followed by four overrelaxation steps as a composite sweep. For each studied value of $\beta$, after discarding 10000 composite sweeps for thermalisation, we record a sample of 600 configurations, separated by 2000 composite sweeps along the Markov chain. Observables are measured as averages over this sample as a function of $\beta$, with the error determined by a bootstrap procedure generally consisting of 500 bootstrap steps.  

\subsection{Monopoles in the Maximal Abelian Gauge}
\label{subsec:TDASU3:abelianproj}
\noindent
Our aim is to analyse the deconfinement phase transition in $SU(3)$ gauge theories using a TDA applied to monopole currents following the lines of Sect.~\ref{sect:u1}. Magnetic monopole configurations appear in non-Abelian gauge theories coupled with an adjoint Higgs field, a classic example being the 't~Hooft-Polyakov monopole in the Georgi-Glashow model~\cite{tHooft:1974kcl,Polyakov:1974ek}. In Ref.~\cite{tHooft:1981bkw}, 't~Hooft proposed that a field transforming in the adjoint representation of the gauge group can act as an effective dynamical Higgs field in $SU(N)$ gauge theories. One can fix the gauge in which such an operator is diagonal with ordered eigenvalues. The diagonal elements of the gauge field define {\em Abelian projected} fields. Abelian magnetic monopoles arise at spacetime points in which the operator used to define the Abelian projection has two consecutive eigenvalues that are equal. For a $SU(N)$ gauge theory, an Abelian projection defines $N-1$ species of monopoles. 't Hooft's proposal was first formulated on the lattice in Refs.~\cite{Kronfeld:1987ri,Kronfeld:1987vd}, with the 't Hooft field tensor corresponding to the $N - 1$ Abelian fields constructed explicitly in~\cite{DelDebbio:2002nb}.    

The original proposal by 't Hooft did not specify the potential role of specific Abelian projections. A kinematically relevant gauge fixing is the Maximal Abelian Gauge (MAG), where the adjoint operator is chosen in such a way that the off-diagonal components of the fields are minimized. In Ref.~\cite{Bonati:2010sp}, it has been shown that the MAG is a convenient gauge for detecting magnetic monopole singularities. Due to its phenomenological relevance (see, e.g.,~\cite{Chernodub:1997ay} for an early review), we choose the MAG for our formulation of the monopole current observables, following the approach of Ref.~\cite{Bonati:2013bga}. 

For the $SU(3)$ lattice Yang-Mills theory, we define the adjoint operator $\tilde X(i)$ as
\begin{equation}
\tilde X(i) = \sum_\mu \left[ U_\mu(i) \tilde\lambda U^\dagger_\mu(i)
+U^\dagger_\mu(i-\hat{\mu}) \tilde\lambda U_\mu(i-\hat{\mu})\right] \ , 
\qquad \tilde \lambda = {\rm diag} (1,0,-1) \ .  
\end{equation}
The MAG is defined as the gauge in which $\tilde X(i)$ is diagonal. This gauge choice is equivalent to requiring that the operator 
\begin{eqnarray}
\tilde F_{\rm MAG} (U,g ) = \sum_{\mu,n} 
\mathrm{Tr} \left( g(n) U_\mu(n) g^{\dagger}  (n + \hat{\mu})  \tilde\lambda g (n + \hat{\mu}) U^{\dagger}_\mu(n) g^\dagger(n) \, \tilde\lambda \right) 
\end{eqnarray}
is maximized over $g$, i.e., the gauge fixing transformation $\{ g \}$ can be derived from the condition
\begin{eqnarray}
\{ \tilde g  \}  = \mathop{\rm argmax}_{\{ g \} } \tilde F _{\rm MAG} ( U  , g ) 
\end{eqnarray}
In the MAG, the diagonal elements of the link matrices $\tilde U _{i i}$ read
\begin{eqnarray}
\tilde U _{i i} = r_i e^{i \varphi_i} \ , \qquad \sum_i \varphi_i = 2 \pi n + \delta \varphi \ ,
\end{eqnarray}
where the last condition, which violates perfect Abelianity of the links, is due to residual off-diagonal elements. Angle variables $\phi_i$ are then defined through the redistribution of the excess phase as
\begin{eqnarray}
\phi_i = \varphi_i - \delta  \varphi\frac{ \left| \tilde U_{ii} \right |^{-1}}{\sum_j \left| \tilde U_{jj}\right| ^{-1}} \ .
\end{eqnarray}
We now set $\theta_1 = \phi_1$ and $\theta_2 = - \phi_3$, and use the DeGrand and Toussaint prescription~\cite{PhysRevD.22.2478} for identifying the monopoles associated to each Abelian field. 		

\begin{figure}[t]
\begin{center}
\includegraphics[width=0.43\textwidth]{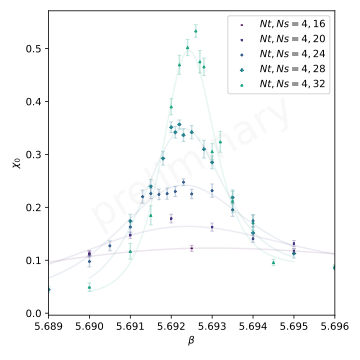}
\hspace{0.1\textwidth}
\includegraphics[width=0.43\textwidth]{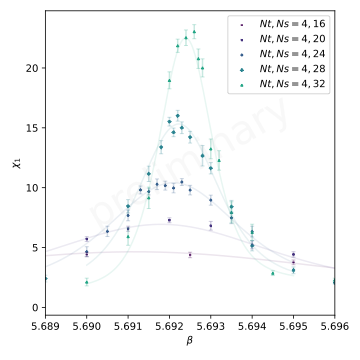}
\end{center}
\caption{Behaviour of $\chi_0$ (left) and $\chi_1$ (right) in $SU(3)$ Lattice Gauge Theory as a function of the coupling $\beta$ at $N_t = 4$ for the indicated values of the lattice sizes. Continuous curves are obtained with a reweighting procedure.\label{fig:sect4:su3chi}}
\end{figure}

\subsection{Numerical results}
\label{subsec:TDASU3:results}
\noindent
We have investigated the behaviour of the quantities $b_0$ and $b_1$, defined in Eqs.~(\ref{eq:betti-number-observables}) and of their susceptibilties $\chi_0$ and $\chi_1$ (see Eqs.~(\ref{eq:betti-number-susceptibilities})) for $SU(3)$ Lattice Gauge Theory across the deconfinement phase transition at $N_t = 4$ and at spacial lattice sizes $N_s = 16, 20, 24, 28, 32$. Our TDA observables have been defined considering the joint set of Abelian monopole currents corresponding to the angles $\theta_1$ and $\theta_2$ extracted after MAG fixing and Abelian projection, as described in the previous subsection. We have verified that considering the network of each Abelian monopole current separately we obtain compatible results. No statistically significant difference have been observed between the networks of the two Abelian monopole currents.
\begin{figure}[t]
\begin{center}
\includegraphics[width=0.43\textwidth]{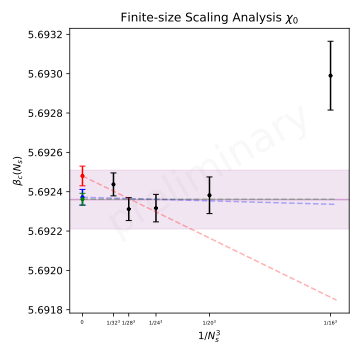}
\hspace{0.1\textwidth}
\includegraphics[width=0.43\textwidth]{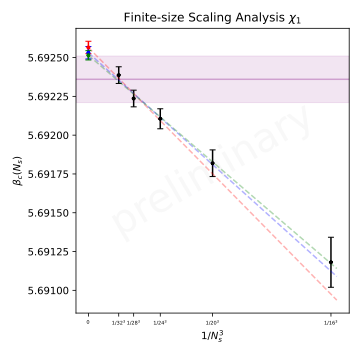}
\end{center}
\caption{Finite-size scaling analysis for the position of the peaks of the susceptibilities  $\chi_0$ (left) and $\chi_1$ (right) in SU(3) Lattice Gauge Theory as a function of the inverse spatial volume at $N_t = 4$. Fits results are indicated for various choices of included lattices as the intercept with the vertical axis. The horizontal band represents the literature value of the critical coupling in the thermodynamic limit (see text for details). \label{fig:sect4:fsssu3}}
\end{figure}

We report the behaviour of $\chi_0$ and $\chi_1$ in Fig~\ref{fig:sect4:su3chi}. Like in the Compact $U(1)$ case, in both quantities we observe a peak of increasing height and shrinking width as the volume increases. The quantities $b_0$ and $b_1$ also behave similarly to the Compact $U(1)$ case.

To confirm that these peaks scale with the expected behaviour at a first order phase transition, we fit their position $\beta_c(N_s)$ with the finite size linear ans\"atz
\begin{equation}
\beta_c(N_s) = \beta_c+ \frac{a}{N_s^3} \ , 
\end{equation}
where $\beta_c$ is the deconfinement critical $\beta$ at $N_t = 4$ and $a$ parametrises the finite-size corrections. Our results are presented in Fig.~\ref{fig:sect4:fsssu3}. Different fits excluding in turn smaller lattice sizes are shown. We note a very good quality of the fits and an excellent agreement with the literature value $\beta_c = 5.69236(15)$ (see, e.g., Ref.~\cite{Lucini:2003zr}), with a significant reduction of the error bars for our indicated fit results. 

Our findings suggest that the features exposed by a TDA analysis of Abelian monopole currents defined in the MAG capture the salient physical properties of the deconfinement phase transition. 

\section{Conclusions}
\label{sect:conclusions}
\noindent
In this contribution, we have developed a TDA approach to the study of monopole current networks in Abelian and non-Abelian gauge theories. We have shown that our observables precisely capture the quantitative features of the deconfinement phase transitions in Compact U(1) Lattice Gauge Theory at zero temperature and in $SU(3)$ Yang-Mills, using in the latter case four lattice sites in the time direction. For $SU(3)$ Yang-Mills, our approach provides noticeably better precision than the conventional analysis based on the study of the Polyakov loop and of its susceptibility. This gives reassuring indication that our observables are coupled to the degrees of freedom that are relevant for the deconfinement phase transition. If this is indeed the case, TDA observables are expected to be sensitive to the transition (or cross-over) from the stringy fluid phase to the deconfinement phase in full QCD. In the next stages of our investigation, we will first extend our $SU(3)$ Yang-Mills study to finer lattice spacings, in order to ascertain that our approach is not significantly affected by lattice artefacts, and then apply our methodology to full QCD, with the goal of verifying the existence of the stringy fluid regime.\\
\noindent
{\bf Note Added:} After this work was presented at the Lattice conference, the publications~\cite{Mickley:2024vkm}~and~\cite{Spitz:2024bqh} appeared, reporting related complementary investigations of the phase structure in QCD and in $SU(3)$ Yang-Mills, with the same goal to understand whether a stringy fluid regime exists.  

\acknowledgments
\noindent
XC was supported by the Additional Funding Programme for Mathematical Sciences, delivered by EPSRC (EP/V521917/1) and the Heilbronn Institute for Mathematical Research. JG was supported by EPSRC grant EP/R018472/1 through the Centre for TDA and the Erlangen Hub for AI through EPSRC grant EP/Y028872/1. The work of BL was partly supported by the EPSRC ExCALIBUR ExaTEPP project EP/X017168/1 and by the STFC Consolidated Grants No. ST/T000813/1 and ST/X000648/1. Numerical simulations have been performed on the Swansea SUNBIRD cluster, part of the Supercomputing Wales project. Supercomputing Wales is part funded by the European Regional Development Fund (ERDF) via Welsh Government. 

\noindent
{\bf Data and Code} --
The data and code used in this manuscript are avaialble from the authors.

\noindent
{\bf Open Access Statement} -- For the purpose of open access, the authors have applied a Creative Commons Attribution (CC BY) licence to any Author Accepted Manuscript version arising.

\bibliographystyle{JHEP}
\bibliography{ref}

\providecommand{\href}[2]{#2}\begingroup\raggedright\begin{thebibliography}{10}

\bibitem{Aarts:2023vsf}
G.~Aarts et~al., \emph{{Phase Transitions in Particle Physics}: {Results and
  Perspectives from Lattice Quantum Chromo-Dynamics}},
  \href{https://doi.org/10.1016/j.ppnp.2023.104070}{\emph{Prog. Part. Nucl.
  Phys.} {\bfseries 133} (2023) 104070}
  [\href{https://arxiv.org/abs/2301.04382}{{\ttfamily 2301.04382}}].

\bibitem{Glozman:2024xll}
L.~Glozman, \emph{Chiral spin symmetry and hot/dense qcd},
  \href{https://doi.org/https://doi.org/10.1016/j.ppnp.2023.104049}{\emph{Progress
  in Particle and Nuclear Physics} {\bfseries 131} (2023) 104049}
  [\href{https://arxiv.org/abs/2404.02606}{{\ttfamily 2404.02606}}].

\bibitem{Hanada:2023krw}
M.~Hanada, H.~Ohata, H.~Shimada and H.~Watanabe, \emph{{A New Perspective on
  Thermal Transition in QCD}},
  \href{https://doi.org/10.1093/ptep/ptae044}{\emph{PTEP} {\bfseries 2024}
  (2024) 041B02} [\href{https://arxiv.org/abs/2310.01940}{{\ttfamily
  2310.01940}}].

\bibitem{tHooft:1977nqb}
G.~'t~Hooft, \emph{{On the Phase Transition Towards Permanent Quark
  Confinement}},
  \href{https://doi.org/10.1016/0550-3213(78)90153-0}{\emph{Nucl. Phys. B}
  {\bfseries 138} (1978) 1}.

\bibitem{tHooft:1981bkw}
G.~'t~Hooft, \emph{{Topology of the Gauge Condition and New Confinement Phases
  in Nonabelian Gauge Theories}},
  \href{https://doi.org/10.1016/0550-3213(81)90442-9}{\emph{Nucl. Phys. B}
  {\bfseries 190} (1981) 455}.

\bibitem{DiGiacomo:1999fb}
A.~Di~Giacomo, B.~Lucini, L.~Montesi and G.~Paffuti, \emph{{Color confinement
  and dual superconductivity of the vacuum. 2.}},
  \href{https://doi.org/10.1103/PhysRevD.61.034504}{\emph{Phys. Rev. D}
  {\bfseries 61} (2000) 034504}
  [\href{https://arxiv.org/abs/hep-lat/9906025}{{\ttfamily hep-lat/9906025}}].

\bibitem{DiGiacomo:1999yas}
A.~Di~Giacomo, B.~Lucini, L.~Montesi and G.~Paffuti, \emph{{Color confinement
  and dual superconductivity of the vacuum. 1.}},
  \href{https://doi.org/10.1103/PhysRevD.61.034503}{\emph{Phys. Rev. D}
  {\bfseries 61} (2000) 034503}
  [\href{https://arxiv.org/abs/hep-lat/9906024}{{\ttfamily hep-lat/9906024}}].

\bibitem{Carmona:2001ja}
J.M.~Carmona, M.~D'Elia, A.~Di~Giacomo, B.~Lucini and G.~Paffuti, \emph{{Color
  confinement and dual superconductivity of the vacuum. 3.}},
  \href{https://doi.org/10.1103/PhysRevD.64.114507}{\emph{Phys. Rev. D}
  {\bfseries 64} (2001) 114507}
  [\href{https://arxiv.org/abs/hep-lat/0103005}{{\ttfamily hep-lat/0103005}}].

\bibitem{Carmona:2002ty}
J.M.~Carmona, M.~D'Elia, L.~Del~Debbio, A.~Di~Giacomo, B.~Lucini and
  G.~Paffuti, \emph{{Color confinement and dual superconductivity in full
  QCD}}, \href{https://doi.org/10.1103/PhysRevD.66.011503}{\emph{Phys. Rev. D}
  {\bfseries 66} (2002) 011503}
  [\href{https://arxiv.org/abs/hep-lat/0205025}{{\ttfamily hep-lat/0205025}}].

\bibitem{Carmona:2002ye}
J.M.~Carmona, M.~D'Elia, L.~Del~Debbio, A.~Di~Giacomo, B.~Lucini, G.~Paffuti
  et~al., \emph{{Color confinement and dual superconductivity in unquenched
  QCD}}, \href{https://doi.org/10.1016/S0375-9474(02)01533-6}{\emph{Nucl. Phys.
  A} {\bfseries 715} (2003) 883}
  [\href{https://arxiv.org/abs/hep-lat/0209080}{{\ttfamily hep-lat/0209080}}].

\bibitem{DElia:2005sfk}
M.~D'Elia, A.~Di~Giacomo, B.~Lucini, G.~Paffuti and C.~Pica, \emph{{Color
  confinement and dual superconductivity of the vacuum. IV.}},
  \href{https://doi.org/10.1103/PhysRevD.71.114502}{\emph{Phys. Rev. D}
  {\bfseries 71} (2005) 114502}
  [\href{https://arxiv.org/abs/hep-lat/0503035}{{\ttfamily hep-lat/0503035}}].

\bibitem{DelDebbio:2000cx}
L.~Del~Debbio, A.~Di~Giacomo and B.~Lucini, \emph{{Vortices, monopoles and
  confinement}},
  \href{https://doi.org/10.1016/S0550-3213(00)00651-9}{\emph{Nucl. Phys. B}
  {\bfseries 594} (2001) 287}
  [\href{https://arxiv.org/abs/hep-lat/0006028}{{\ttfamily hep-lat/0006028}}].

\bibitem{DelDebbio:2000cb}
L.~Del~Debbio, A.~Di~Giacomo and B.~Lucini, \emph{{Monopoles, vortices and
  confinement in SU(3) gauge theory}},
  \href{https://doi.org/10.1016/S0370-2693(01)00091-0}{\emph{Phys. Lett. B}
  {\bfseries 500} (2001) 326}
  [\href{https://arxiv.org/abs/hep-lat/0011048}{{\ttfamily hep-lat/0011048}}].

\bibitem{Greensite:2008ss}
J.~Greensite and B.~Lucini, \emph{{Is Confinement a Phase of Broken Dual Gauge
  Symmetry?}}, \href{https://doi.org/10.1103/PhysRevD.78.085004}{\emph{Phys.
  Rev. D} {\bfseries 78} (2008) 085004}
  [\href{https://arxiv.org/abs/0806.2117}{{\ttfamily 0806.2117}}].

\bibitem{PhysRevResearch.2.043308}
B.~Olsthoorn, J.~Hellsvik and A.V.~Balatsky, \emph{Finding hidden order in spin
  models with persistent homology},
  \href{https://doi.org/10.1103/PhysRevResearch.2.043308}{\emph{Phys. Rev.
  Res.} {\bfseries 2} (2020) 043308}.

\bibitem{PhysRevB.104.104426}
A.~Cole, G.J.~Loges and G.~Shiu, \emph{Quantitative and interpretable order
  parameters for phase transitions from persistent homology},
  \href{https://doi.org/10.1103/PhysRevB.104.104426}{\emph{Phys. Rev. B}
  {\bfseries 104} (2021) 104426}.

\bibitem{PhysRevD.107.034501}
N.~Sale, B.~Lucini and J.~Giansiracusa, \emph{Probing center vortices and
  deconfinement in su(2) lattice gauge theory with persistent homology},
  \href{https://doi.org/10.1103/PhysRevD.107.034501}{\emph{Phys. Rev. D}
  {\bfseries 107} (2023) 034501}.

\bibitem{PhysRevB.104.235146}
A.~Tirelli and N.C.~Costa, \emph{Learning quantum phase transitions through
  topological data analysis},
  \href{https://doi.org/10.1103/PhysRevB.104.235146}{\emph{Phys. Rev. B}
  {\bfseries 104} (2021) 235146}.

\bibitem{PhysRevE.105.024121}
N.~Sale, J.~Giansiracusa and B.~Lucini, \emph{Quantitative analysis of phase
  transitions in two-dimensional $xy$ models using persistent homology},
  \href{https://doi.org/10.1103/PhysRevE.105.024121}{\emph{Phys. Rev. E}
  {\bfseries 105} (2022) 024121}.

\bibitem{PhysRevB.106.085111}
D.~Sehayek and R.G.~Melko, \emph{Persistent homology of $\mathbb{Z}_{2}$ gauge
  theories}, \href{https://doi.org/10.1103/PhysRevB.106.085111}{\emph{Phys.
  Rev. B} {\bfseries 106} (2022) 085111}.

\bibitem{hirakida2020persistent}
T.~Hirakida, K.~Kashiwa, J.~Sugano, J.~Takahashi, H.~Kouno and M.~Yahiro,
  \emph{Persistent homology analysis of deconfinement transition in effective
  polyakov-line model},
  \href{https://doi.org/10.1142/S0217751X20500499}{\emph{International Journal
  of Modern Physics A} {\bfseries 35} (2020) 2050049}.

\bibitem{PhysRevD.108.056016}
D.~Spitz, K.~Boguslavski and J.~Berges, \emph{Probing universal dynamics with
  topological data analysis in a gluonic plasma},
  \href{https://doi.org/10.1103/PhysRevD.108.056016}{\emph{Phys. Rev. D}
  {\bfseries 108} (2023) 056016}.

\bibitem{crean2024tda}
X.~Crean, J.~Giansiracusa and B.~Lucini, \emph{Topological data analysis of
  monopole current networks in $ u (1) $ lattice gauge theory}, {\emph{SciPost
  Physics} {\bfseries 17} (2024) 100}.

\bibitem{topology-hypothesis1}
L.~Caiani, L.~Casetti, C.~Clementi and M.~Pettini, \emph{Geometry of dynamics,
  lyapunov exponents, and phase transitions},
  \href{https://doi.org/10.1103/PhysRevLett.79.4361}{\emph{Phys. Rev. Lett.}
  {\bfseries 79} (1997) 4361}.

\bibitem{topology-hypothesis2}
M.~Kastner, \emph{Phase transitions and configuration space topology},
  \href{https://doi.org/10.1103/RevModPhys.80.167}{\emph{Rev. Mod. Phys.}
  {\bfseries 80} (2008) 167}.

\bibitem{topology-hypothesis-theorem}
R.~Franzosi and M.~Pettini, \emph{Theorem on the origin of phase transitions},
  \href{https://doi.org/10.1103/PhysRevLett.92.060601}{\emph{Phys. Rev. Lett.}
  {\bfseries 92} (2004) 060601}.

\bibitem{topology-hypothesis-with-PH}
I.~Donato, M.~Gori, M.~Pettini, G.~Petri, S.~De~Nigris, R.~Franzosi et~al.,
  \emph{Persistent homology analysis of phase transitions},
  \href{https://doi.org/10.1103/PhysRevE.93.052138}{\emph{Phys. Rev. E}
  {\bfseries 93} (2016) 052138}.

\bibitem{PhysRevE.98.012318}
L.~Speidel, H.A.~Harrington, S.J.~Chapman and M.A.~Porter, \emph{Topological
  data analysis of continuum percolation with disks},
  \href{https://doi.org/10.1103/PhysRevE.98.012318}{\emph{Phys. Rev. E}
  {\bfseries 98} (2018) 012318}.

\bibitem{PhysRevE.103.052127}
Q.H.~Tran, M.~Chen and Y.~Hasegawa, \emph{Topological persistence machine of
  phase transitions},
  \href{https://doi.org/10.1103/PhysRevE.103.052127}{\emph{Phys. Rev. E}
  {\bfseries 103} (2021) 052127}.

\bibitem{sym14091783}
K.~Kashiwa, T.~Hirakida and H.~Kouno, \emph{Persistent homology analysis for
  dense qcd effective model with heavy quarks},
  \href{https://doi.org/10.3390/sym14091783}{\emph{Symmetry} {\bfseries 14}
  (2022) }.

\bibitem{TDA-observable4}
D.~Spitz, J.M.~Urban and J.M.~Pawlowski, \emph{Confinement in non-abelian
  lattice gauge theory via persistent homology},
  \href{https://doi.org/10.1103/PhysRevD.107.034506}{\emph{Phys. Rev. D}
  {\bfseries 107} (2023) 034506}.

\bibitem{gudhi}
\emph{GUDHI User and Reference Manual}, GUDHI Editorial Board, 3.10.1~ed.
  (2024).

\bibitem{Bauer-Lesnick}
U.~Bauer and M.~Lesnick, \emph{Induced matchings and the algebraic stability of
  persistence barcodes}, {\emph{Journal of Computational Geometry} {\bfseries
  6} (2015) 162}.

\bibitem{Nanda-vectorisation-survey}
D.~Ali, A.~Asaad, M.-J.~Jimenez, V.~Nanda, E.~Paluzo-Hidalgo and
  M.~Soriano-Trigueros, \emph{A survey of vectorization methods in topological
  data analysis}, \href{https://doi.org/10.1109/TPAMI.2023.3308391}{\emph{IEEE
  Transactions on Pattern Analysis and Machine Intelligence} {\bfseries 45}
  (2023) 14069}.

\bibitem{Adams-persistence-images}
H.~Adams, T.~Emerson, M.~Kirby, R.~Neville, C.~Peterson, P.~Shipman et~al.,
  \emph{Persistence images: A stable vector representation of persistent
  homology}, {\emph{Journal of Machine Learning Research} {\bfseries 18} (2017)
  1}.

\bibitem{PhysRevD.22.2478}
T.A.~DeGrand and D.~Toussaint, \emph{Topological excitations and monte carlo
  simulation of abelian gauge theory},
  \href{https://doi.org/10.1103/PhysRevD.22.2478}{\emph{Phys. Rev. D}
  {\bfseries 22} (1980) 2478}.

\bibitem{Creutz:1979zg}
M.~Creutz, L.~Jacobs and C.~Rebbi, \emph{{Monte Carlo Study of Abelian Lattice
  Gauge Theories}}, \href{https://doi.org/10.1103/PhysRevD.20.1915}{\emph{Phys.
  Rev. D} {\bfseries 20} (1979) 1915}.

\bibitem{Lautrup:1980xr}
B.E.~Lautrup and M.~Nauenberg, \emph{{Phase Transition in Four-Dimensional
  Compact QED}},
  \href{https://doi.org/10.1016/0370-2693(80)90400-1}{\emph{Phys. Lett. B}
  {\bfseries 95} (1980) 63}.

\bibitem{Bhanot:1981tn}
G.~Bhanot, \emph{{Compact {QED} With an Extended Lattice Action}},
  \href{https://doi.org/10.1016/0550-3213(82)90382-0}{\emph{Nucl. Phys. B}
  {\bfseries 205} (1982) 168}.

\bibitem{Jersak:1983yz}
J.~Jersak, T.~Neuhaus and P.M.~Zerwas, \emph{{U(1) Lattice Gauge Theory Near
  the Phase Transition}},
  \href{https://doi.org/10.1016/0370-2693(83)90115-6}{\emph{Phys. Lett. B}
  {\bfseries 133} (1983) 103}.

\bibitem{Barber:1984ak}
J.S.~Barber, R.E.~Shrock and R.~Schrader, \emph{{A Study of $d=4$ U(1) Lattice
  Gauge Theory With Monopoles Removed}},
  \href{https://doi.org/10.1016/0370-2693(85)91174-8}{\emph{Phys. Lett. B}
  {\bfseries 152} (1985) 221}.

\bibitem{Evertz:1984pn}
H.G.~Evertz, J.~Jersak, T.~Neuhaus and P.M.~Zerwas, \emph{{Tricritical Point in
  Lattice {QED}}},
  \href{https://doi.org/10.1016/0550-3213(85)90262-7}{\emph{Nucl. Phys. B}
  {\bfseries 251} (1985) 279}.

\bibitem{Barber:1985at}
J.S.~Barber and R.E.~Shrock, \emph{{Dynamical Shifting of the
  Confinement-Deconfinement Phase Transition in 4D U(1) Lattice Gauge Theory}},
  \href{https://doi.org/10.1016/0550-3213(85)90361-X}{\emph{Nucl. Phys. B}
  {\bfseries 257} (1985) 515}.

\bibitem{GROSCH1985171}
V.~Grösch, K.~Jansen, J.~Jersák, C.~Lang, T.~Neuhaus and C.~Rebbi,
  \emph{Monopoles and dirac sheets in compact u(1) lattice gauge theory},
  \href{https://doi.org/10.1016/0370-2693(85)91081-0}{\emph{Physics Letters B}
  {\bfseries 162} (1985) 171}.

\bibitem{Lang:1986yd}
C.B.~Lang, \emph{{Renormalization study of compact U(1) lattice gauge theory}},
  \href{https://doi.org/10.1016/0550-3213(87)90147-7}{\emph{Nucl. Phys. B}
  {\bfseries 280} (1987) 255}.

\bibitem{DelDebbio:1994sx}
L.~Del~Debbio, A.~Di~Giacomo and G.~Paffuti, \emph{{Detecting dual
  superconductivity in the ground state of gauge theory}},
  \href{https://doi.org/10.1016/0370-2693(95)00266-N}{\emph{Phys. Lett. B}
  {\bfseries 349} (1995) 513}
  [\href{https://arxiv.org/abs/hep-lat/9403013}{{\ttfamily hep-lat/9403013}}].

\bibitem{Lang:1994ri}
C.B.~Lang and T.~Neuhaus, \emph{{Compact U(1) gauge theory on lattices with
  trivial homotopy group}},
  \href{https://doi.org/10.1016/0550-3213(94)90100-7}{\emph{Nucl. Phys. B}
  {\bfseries 431} (1994) 119}
  [\href{https://arxiv.org/abs/hep-lat/9407005}{{\ttfamily hep-lat/9407005}}].

\bibitem{Kerler:1994qc}
W.~Kerler, C.~Rebbi and A.~Weber, \emph{{Phase structure and monopoles in U(1)
  gauge theory}}, \href{https://doi.org/10.1103/PhysRevD.50.6984}{\emph{Phys.
  Rev. D} {\bfseries 50} (1994) 6984}
  [\href{https://arxiv.org/abs/hep-lat/9403025}{{\ttfamily hep-lat/9403025}}].

\bibitem{Kerler:1995va}
W.~Kerler, C.~Rebbi and A.~Weber, \emph{{Monopole currents and Dirac sheets in
  U(1) lattice gauge theory}},
  \href{https://doi.org/10.1016/0370-2693(95)00188-Q}{\emph{Phys. Lett. B}
  {\bfseries 348} (1995) 565}
  [\href{https://arxiv.org/abs/hep-lat/9501023}{{\ttfamily hep-lat/9501023}}].

\bibitem{Kerler:1995nj}
W.~Kerler, C.~Rebbi and A.~Weber, \emph{{Phase transition and dynamical
  parameter method in U(1) gauge theory}},
  \href{https://doi.org/10.1016/0550-3213(95)00239-O}{\emph{Nucl. Phys. B}
  {\bfseries 450} (1995) 452}
  [\href{https://arxiv.org/abs/hep-lat/9503021}{{\ttfamily hep-lat/9503021}}].

\bibitem{Jersak:1996mn}
J.~Jersak, C.B.~Lang and T.~Neuhaus, \emph{{NonGaussian fixed point in
  four-dimensional pure compact U(1) gauge theory on the lattice}},
  \href{https://doi.org/10.1103/PhysRevLett.77.1933}{\emph{Phys. Rev. Lett.}
  {\bfseries 77} (1996) 1933}
  [\href{https://arxiv.org/abs/hep-lat/9606010}{{\ttfamily hep-lat/9606010}}].

\bibitem{Jersak:1996mj}
J.~Jersak, C.B.~Lang and T.~Neuhaus, \emph{{Four-dimensional pure compact U(1)
  gauge theory on a spherical lattice}},
  \href{https://doi.org/10.1103/PhysRevD.54.6909}{\emph{Phys. Rev. D}
  {\bfseries 54} (1996) 6909}
  [\href{https://arxiv.org/abs/hep-lat/9606013}{{\ttfamily hep-lat/9606013}}].

\bibitem{Kerler:1996sf}
W.~Kerler, C.~Rebbi and A.~Weber, \emph{{Order parameters and boundary effects
  in U(1) lattice gauge theory}},
  \href{https://doi.org/10.1016/0370-2693(96)00498-4}{\emph{Phys. Lett. B}
  {\bfseries 380} (1996) 346}
  [\href{https://arxiv.org/abs/hep-lat/9601002}{{\ttfamily hep-lat/9601002}}].

\bibitem{Kerler:1996cr}
W.~Kerler, C.~Rebbi and A.~Weber, \emph{{Critical properties and monopoles in
  U(1) lattice gauge theory}},
  \href{https://doi.org/10.1016/S0370-2693(96)01564-X}{\emph{Phys. Lett. B}
  {\bfseries 392} (1997) 438}
  [\href{https://arxiv.org/abs/hep-lat/9612001}{{\ttfamily hep-lat/9612001}}].

\bibitem{Campos:1997br}
I.~Campos, A.~Cruz and A.~Tarancon, \emph{{First order signatures in 4-D pure
  compact U(1) gauge theory with toroidal and spherical topologies}},
  \href{https://doi.org/10.1016/S0370-2693(98)00208-1}{\emph{Phys. Lett. B}
  {\bfseries 424} (1998) 328}
  [\href{https://arxiv.org/abs/hep-lat/9711045}{{\ttfamily hep-lat/9711045}}].

\bibitem{Campos:1998jp}
I.~Campos, A.~Cruz and A.~Tarancon, \emph{{A Study of the phase transition in
  4-D pure compact U(1) LGT on toroidal and spherical lattices}},
  \href{https://doi.org/10.1016/S0550-3213(98)00452-0}{\emph{Nucl. Phys. B}
  {\bfseries 528} (1998) 325}
  [\href{https://arxiv.org/abs/hep-lat/9803007}{{\ttfamily hep-lat/9803007}}].

\bibitem{Vettorazzo:2003fg}
M.~Vettorazzo and P.~de~Forcrand, \emph{{Electromagnetic fluxes, monopoles, and
  the order of the 4-d compact U(1) phase transition}},
  \href{https://doi.org/10.1016/j.nuclphysb.2004.02.038}{\emph{Nucl. Phys. B}
  {\bfseries 686} (2004) 85}
  [\href{https://arxiv.org/abs/hep-lat/0311006}{{\ttfamily hep-lat/0311006}}].

\bibitem{DiCairano:2023kzh}
L.~Di~Cairano, M.~Gori, M.~Sarkis and A.~Tkatchenko, \emph{{Phase Transitions
  in Abelian Lattice Gauge Theory: Production and Dissolution of Monopoles and
  Monopole-Antimonopole Pairs}}, .

\bibitem{Arnold_2001}
G.~Arnold, T.~Lippert, K.~Schilling and T.~Neuhaus, \emph{Finite size scaling
  analysis of compact qed},
  \href{https://doi.org/10.1016/s0920-5632(01)01001-5}{\emph{Nuclear Physics B
  - Proceedings Supplements} {\bfseries 94} (2001) 651–656}.

\bibitem{langfeld2016efficient}
K.~Langfeld, B.~Lucini, R.~Pellegrini and A.~Rago, \emph{An efficient algorithm
  for numerical computations of continuous densities of states},
  \href{https://doi.org/10.1140/epjc/s10052-016-4142-5}{\emph{Th European
  Physical Journal C} {\bfseries 76} (2016) 306}.

\bibitem{ZENODO2}
X.~Crean, J.~Giansiracusa and B.~Lucini, \emph{{Topological Data Analysis of
  Monopoles in $U(1)$ Lattice Gauge Theory---Monte Carlo and analysis code
  release}},  2024.
\newblock 10.5281/zenodo.10806185.

\bibitem{Bazavov:2005zy}
A.~Bazavov and B.A.~Berg, \emph{{Heat bath efficiency with metropolis-type
  updating}}, \href{https://doi.org/10.1103/PhysRevD.71.114506}{\emph{Phys.
  Rev. D} {\bfseries 71} (2005) 114506}
  [\href{https://arxiv.org/abs/hep-lat/0503006}{{\ttfamily hep-lat/0503006}}].

\bibitem{ferrenberg1989optimized}
A.M.~Ferrenberg and R.H.~Swendsen, \emph{Optimized monte carlo data analysis},
  \href{https://doi.org/10.1103/PhysRevLett.63.1195}{\emph{Phys. Rev. Lett.}
  {\bfseries 63} (1989) 1195}.

\bibitem{tHooft:1974kcl}
G.~'t~Hooft, \emph{{Magnetic Monopoles in Unified Gauge Theories}},
  \href{https://doi.org/10.1016/0550-3213(74)90486-6}{\emph{Nucl. Phys. B}
  {\bfseries 79} (1974) 276}.

\bibitem{Polyakov:1974ek}
A.M.~Polyakov, \emph{{Particle Spectrum in Quantum Field Theory}}, {\emph{JETP
  Lett.} {\bfseries 20} (1974) 194}.

\bibitem{Kronfeld:1987ri}
A.S.~Kronfeld, M.L.~Laursen, G.~Schierholz and U.J.~Wiese, \emph{{Monopole
  Condensation and Color Confinement}},
  \href{https://doi.org/10.1016/0370-2693(87)90910-5}{\emph{Phys. Lett. B}
  {\bfseries 198} (1987) 516}.

\bibitem{Kronfeld:1987vd}
A.S.~Kronfeld, G.~Schierholz and U.J.~Wiese, \emph{{Topology and Dynamics of
  the Confinement Mechanism}},
  \href{https://doi.org/10.1016/0550-3213(87)90080-0}{\emph{Nucl. Phys. B}
  {\bfseries 293} (1987) 461}.

\bibitem{DelDebbio:2002nb}
L.~Del~Debbio, A.~Di~Giacomo, B.~Lucini and G.~Paffuti, \emph{{Abelian
  projection in SU(N) gauge theories}},
  \href{https://arxiv.org/abs/hep-lat/0203023}{{\ttfamily hep-lat/0203023}}.

\bibitem{Bonati:2010sp}
C.~Bonati, M.~D'Elia, A.~Di~Giacomo, L.~Lepori and F.~Pucci, \emph{{Non abelian
  Bianchi identities, monopoles and gauge invariance}},
  \href{https://doi.org/10.22323/1.105.0270}{\emph{PoS} {\bfseries LATTICE2010}
  (2010) 270} [\href{https://arxiv.org/abs/1011.0371}{{\ttfamily 1011.0371}}].

\bibitem{Chernodub:1997ay}
M.N.~Chernodub and M.I.~Polikarpov, \emph{{Abelian projections and monopoles}},
   in \emph{{NATO Advanced Study Institute on Confinement, Duality and
  Nonperturbative Aspects of QCD}}, pp.~387--414, 6, 1997
  [\href{https://arxiv.org/abs/hep-th/9710205}{{\ttfamily hep-th/9710205}}].

\bibitem{Bonati:2013bga}
C.~Bonati and M.~D'Elia, \emph{{The Maximal Abelian Gauge in SU(N) gauge
  theories and thermal monopoles for N = 3}},
  \href{https://doi.org/10.1016/j.nuclphysb.2013.10.004}{\emph{Nucl. Phys. B}
  {\bfseries 877} (2013) 233}
  [\href{https://arxiv.org/abs/1308.0302}{{\ttfamily 1308.0302}}].

\bibitem{Lucini:2003zr}
B.~Lucini, M.~Teper and U.~Wenger, \emph{{The High temperature phase transition
  in SU(N) gauge theories}},
  \href{https://doi.org/10.1088/1126-6708/2004/01/061}{\emph{JHEP} {\bfseries
  01} (2004) 061} [\href{https://arxiv.org/abs/hep-lat/0307017}{{\ttfamily
  hep-lat/0307017}}].

\bibitem{Mickley:2024vkm}
J.A.~Mickley, C.~Allton, R.~Bignell and D.B.~Leinweber, \emph{{Centre vortex
  evidence for a second finite-temperature QCD transition}},
  \href{https://arxiv.org/abs/2411.19446}{{\ttfamily 2411.19446}}.

\bibitem{Spitz:2024bqh}
D.~Spitz, J.M.~Urban and J.M.~Pawlowski, \emph{{Topological data analysis of
  the deconfinement transition in SU(3) lattice gauge theory}},
  \href{https://arxiv.org/abs/2412.09112}{{\ttfamily 2412.09112}}.

\end{thebibliography}\endgroup

\end{document}